\documentclass[twocolumn,showpacs,preprintnumbers,amsmath,amssymb]{revtex4}
\usepackage{graphicx}
\usepackage{dcolumn}
\usepackage{bm}

\usepackage{color}

\begin{document}

\title{Back-action amplification
and quantum limits in optomechanical measurements}


\author{P. Verlot}
\author{A. Tavernarakis}
\author{T. Briant}
\author{P.-F. Cohadon}
\author{A. Heidmann}

\affiliation{Laboratoire Kastler Brossel, ENS, UPMC, CNRS; case 74,
4 place Jussieu, 75005 Paris, France}


\begin{abstract}
Optical interferometry is by far the most sensitive displacement
measurement technique available, with sensitivities at the
$10^{-20}\,{\rm m}/\sqrt{\rm Hz}$ level in the large-scale
gravitational-wave interferometers currently in operation. Second
generation interferometers will experience a 10-fold improvement
in sensitivity and be mainly limited by quantum noise, close to
the Standard Quantum Limit (SQL), once considered as the ultimate
displacement sensitivity achievable by interferometry. In this
Letter, we experimentally demonstrate one of the techniques
envisioned to go beyond the SQL:  amplification of a signal by
radiation-pressure back-action in a detuned cavity.
\end{abstract}


\maketitle

Gravitational-wave (GW) astronomy \cite{Bradaschia90,Abramovici92}
is no longer the sole field of application of high-sensitivity
interferometric displacement measurements, now at work as well in
condensed-matter experiments such as single-spin magnetic
resonance force microscopy \cite{Rugar}, persistent-current
detection in superconductors \cite{Anya}, or the quest to quantum
effects in mesoscopic mechanical systems \cite{Schwab,Lehnert1}.
Recent developments have led both future advanced GW
interferometers \cite{AdvLIGO,AdvVirgo} and current
micro-optomechanical systems \cite{Schliesser,Teufel,Anetsberger}
close to the Standard Quantum Limit (SQL)
\cite{Caves81,Jaekel90,Braginsky92}, where quantum fluctuations of
radiation pressure have observable back-action effects upon the
moving mirror and the measurement sensitivity.

Quantum effects of radiation pressure are so weak that they
haven't been experimentally demonstrated yet, though a number of
dedicated experiments are getting closer, either by a combination
of high optical power and ultra low-mass mirrors \cite{Schnabel}
or by a careful examination of optomechanical correlations between
two light beams sent into the same moving mirror cavity
\cite{VerlotPRL09}. As future experiments will be confronted with
the SQL, a number of schemes have been devised to go beyond it,
either by sending squeezed light into the moving mirror cavity
\cite{Jaekel90,Kimble01} or by performing a back-action evading
measurement \cite{Schwab09} with a two-tone drive of the
optomechanical cavity.

Another approach takes advantage of the radiation-pressure
back-action in a moving mirror cavity: for a non-zero cavity
detuning, a small cavity-length variation induces an intracavity
radiation-pressure modulation which drives the mirror into motion.
This may amplify the signal and lead to a sensitivity beyond the
SQL, either in signal-recycled GW interferometers
\cite{Buonanno2001,Buonanno2002} or in a single detuned optical
cavity \cite{Arcizet2006}.  In this Letter, we report the
observation of such an amplification effect by radiation-pressure
back-action and we demonstrate its ability to improve the
sensitivity beyond the SQL.

\begin{figure}
\includegraphics[width=6 cm]{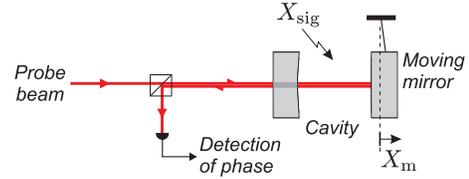}
\caption{Optical measurement of a cavity length variation $X_{\rm
sig}$ through the phase of the reflected beam. Displacements $X_{\rm
m}$ of the moving mirror limit the sensitivity.}
\label{fig:Principle}
\end{figure}

We consider a single-port cavity of length $L$ with a fixed,
partially transmitting front mirror, and a moving, perfectly
reflecting end mirror (see fig. \ref{fig:Principle}). The signal
$X_{\rm{sig}}$ is a cavity-length modulation which is superimposed
to the mirror displacement noise $X_{\rm{m}}$, leading to a
variable cavity detuning $\psi\equiv 2kL [2\pi]$ given by
\begin{equation}
\psi(t) = \overline{\psi}+2k
\left(X_{\rm{m}}(t)+X_{\rm{sig}}(t)\right),\label{eq:detuning}
\end{equation}
where $k$ is the field wavevector and $\overline{\psi}$ the mean
detuning. Incident, intracavity and reflected fields obey the
usual Fabry-Perot cavity relations except for the time dependence
of the cavity detuning $\psi$. In Fourier space, one then gets
linearized input-output relations for the respective annihilation
operators $a^{\rm in}[\Omega]$, $a[\Omega]$, $a^{\rm out}[\Omega]$
of these fields at frequency $\Omega$,
\begin{eqnarray}
(\gamma-i\overline{\psi}-i\Omega\tau)a[\Omega]&=&
\sqrt{2\gamma}a^{\rm in}[\Omega]
+i\overline{a} \psi[\Omega], \label{eq:dyn} \\
a^{\rm out}[\Omega]&=&-a^{\rm in}[\Omega]+
\sqrt{2\gamma}a[\Omega],\label{eq:in-out}
\end{eqnarray}
where $\gamma$ is the damping rate of the cavity, assumed to be
small compared to unity, and $\tau$ is the cavity storage time.

According to (\ref{eq:detuning}) and (\ref{eq:dyn}) the dynamics
of the intracavity field are equally sensitive to the signal
$X_{\rm sig}[\Omega]$ and to the mirror displacements $X_{\rm
m}[\Omega]$. The outgoing field [eq. (\ref{eq:in-out})] then
reflects both the signal and mirror displacement noise, including
radiation-pressure noise.

The radiation-pressure force $F_{\rm{rad}}=2 \hbar k I$ is
proportional to the intracavity intensity $I=a^{\dagger}a$. From
(\ref{eq:detuning}) and (\ref{eq:dyn}), it can be written as the sum
of three terms \cite{Arcizet2006}: the first one related to the
incident field fluctuations $a^{\rm in}[\Omega]$ corresponds to
radiation-pressure quantum noise, the two others related to $X_{\rm
m}$ and $X_{\rm sig}$ are given by
\begin{eqnarray}
F_{\rm rad}^{\left({\rm m}\right)}\left[\Omega\right] &=& -8\hbar
k^2 \overline{I} \frac{\overline \psi}{\Delta} X_{\rm
m}\left[\Omega\right], \label{eq:Frad_m} \\
F_{\rm rad}^{\left({\rm sig}\right)}\left[\Omega\right] &=&
-8\hbar k^2 \overline{I} \frac{\overline \psi}{\Delta} X_{\rm
sig}\left[\Omega\right], \label{eq:Frad_sig}
\end{eqnarray}
where $\Delta = \left(\gamma-i\Omega\tau\right)^2+\overline
\psi^2$ and $\overline{I}=|\overline{a}|^2$ is the mean intensity.
The force $F_{\rm {rad}}^{({\rm m})}$ corresponds to the dynamical
back-action which changes the mechanical response of the mirror,
its mechanical susceptibility $\chi$ being modified to an
effective susceptibility $\chi_{\rm eff}$ given by
\begin{equation}
\chi_{\rm eff}^{-1}\left[\Omega\right] =
\chi^{-1}\left[\Omega\right] + 8\hbar k^2 \overline{I}
\frac{\overline \psi}{\Delta}. \label{eq:chieff}
\end{equation}
This force thus leads to additional optical spring and damping on
the mirror in a detuned cavity ($\overline{\psi}\neq 0$), and is
responsible for the radiation-pressure cooling of the mirror in a
red-detuned cavity
\cite{NatureLKB,NatureGigan,NatureHarris,Kippenberg}.

The force of interest in this paper is $F_{\rm rad}^{({\rm
sig})}$: according to (\ref{eq:Frad_sig}) and (\ref{eq:chieff}),
it induces a mirror displacement $X_{\rm m}^{({\rm
sig})}=\chi_{\rm eff}F_{\rm rad}^{({\rm sig})}$ proportional to
the signal $X_{\rm sig}$, leading to a total length variation,
\begin{equation}
X_{\rm m}^{({\rm sig})}[\Omega] + X_{\rm sig}[\Omega] =
\frac{\chi_{\rm eff}[\Omega]}{\chi[\Omega]} X_{\rm
sig}[\Omega].\label{eq:Xm_sig}
\end{equation}
Depending on the ratio between the initial and effective
susceptibilities, one then gets either an amplification or a
deamplification of the signal by the mirror motion.

We now derive the phase of the field reflected by the cavity,
using the usual definition of the phase quadrature $q[\Omega]$ for
any field operator $a$,
\begin{equation}
\left|\overline{a}\right| q[\Omega] = i \left(\overline{a}
a^{\dagger}[\Omega] - \overline{a}^\star a[\Omega]\right).
\label{eq:quad_q}
\end{equation}
Assuming for simplicity that frequencies $\Omega$ of interest are
much smaller than the cavity bandwidth
$\Omega_{\rm{cav}}=\gamma/\tau$, equations (\ref{eq:dyn}) and
(\ref{eq:in-out}) show that the phase $q^{\rm out}$ of the
reflected field simply reproduces the cavity length variations
$X_{\rm m}+X_{\rm sig}$ (including radiation-pressure noise), with
an additional noise term related to the incident phase
fluctuations $q^{\rm in}$,
\begin{equation}
q^{\rm out}[\Omega] = q^{\rm in}[\Omega] + 2\xi \left(X_{\rm
m}[\Omega] + X_{\rm sig}[\Omega] \right), \label{eq:qout}
\end{equation}
where $\xi=4k \gamma \left|\overline a^{\rm in}\right| /
\left(\gamma^2+\overline{\psi}^2\right)$. One finally gets the
spectrum $S_q^{\rm out}[\Omega]$ of the measured phase quadrature
as
\begin{equation}
\frac{S_q^{\rm out}}{4\xi^2} = \frac1{4\xi^2} + \hbar^2 \xi^2
\left|\chi_{\rm eff}\right|^2 + \left|\frac{\chi_{\rm
eff}}{\chi}\right|^2 S_x^{\rm sig},\label{eq:sensibilite}
\end{equation}
where $S_x^{\rm sig}[\Omega]$ is the spectrum of the signal
$X_{\rm sig}$. First two terms in eq. (\ref{eq:sensibilite}) are
the usual quantum shot and radiation-pressure noises: they exactly
correspond to the ones obtained for a resonant cavity with a
mirror having a mechanical susceptibility $\chi_{\rm eff}$. Their
sum can be rewritten as $\left|\hbar \chi_{\rm eff}\right|
\frac{\zeta^{-1}+\zeta}{2}$, which only depends on the
dimensionless optomechanical parameter $\zeta =
2\hbar\xi^2\left|\chi_{\rm {eff}}\right|$. At any frequency
$\Omega$, the sum is minimal and equal to the standard quantum
limit $\left|\hbar \chi_{\rm eff}[\Omega]\right|$ when
$\zeta[\Omega]=1$.

Last term in eq. (\ref{eq:sensibilite}) reflects the signal, but
with an amplification factor $\left|\chi_{\rm eff}/\chi\right|^2$
similar to the one already found in eq. (\ref{eq:Xm_sig}). In
absence of dynamical radiation-pressure effects ($X_{\rm m}^{({\rm
sig})}=0$) as in the case of a resonant cavity, this factor simply
disappears and the second term in eq. (\ref{eq:sensibilite})
reduces to $S_x^{\rm sig}$. It is then clear that dynamical
back-action not only does change the mechanical behavior of the
moving mirror from $\chi$ to $\chi_{\rm eff}$, but also enables an
amplification of the signal proportional to the factor
$\left|\chi_{\rm eff}/\chi\right|^2$. Equation
(\ref{eq:sensibilite}) therefore shows that a high amplification
factor together with an optomechanical parameter $\zeta\simeq 1$
may afford a significant increase of the measurement sensitivity
beyond the SQL $\left|\hbar \chi_{\rm eff}\right|$.

\begin{figure}
\includegraphics[width=8.6 cm]{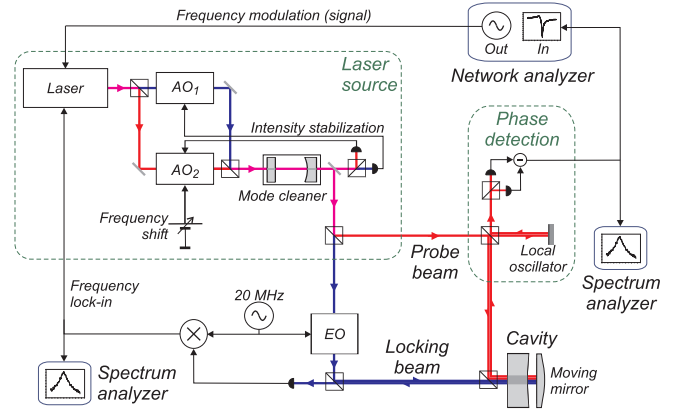}
\caption{Experimental setup. A laser source provides two beams, a
locking beam used to set the laser frequency at the optical
resonance of the cavity via a Pound-Drever-Hall technique, and a
probe beam which can be detuned by an acousto-optic modulator (AO).
The measured signal is provided by a network analyzer which
modulates the laser frequency to induce an optical length modulation
of the cavity. The phase of the reflected probe beam is monitored
with a homodyne detection and the result is sent both to the network
analyzer and to a spectrum analyzer. For simplicity, most polarizing
elements are not shown.} \label{fig:Setup}
\end{figure}

Our experimental setup (figure \ref{fig:Setup}) is based on a
single-ended optical cavity, with a 1-inch fused silica cylindrical
input mirror. The end mirror is coated on a plano-convex 34-mm
diameter and 2.5-mm thick substrate, which exhibits gaussian
internal vibration modes \cite{PRA_modes_gaussiens}. In the
following we only consider frequencies $\Omega$ close to a single
mechanical resonance of this moving mirror, so that its motion can
be considered as the one of a harmonic oscillator characterized by a
lorentzian susceptibility,
\begin{equation}
\chi\left[\Omega\right] = \frac1{M\left(\Omega_{\rm M}^2 - \Omega^2
- i\Omega_{\rm M}\Omega/Q\right)}, \label{eq:chi}
\end{equation}
with the following characteristics, deduced from the thermal noise
spectrum observed at room temperature: resonance frequency
$\Omega_{\rm M}/2\pi\simeq 1128.5\,{\rm kHz}$, mass $M=72\,{\rm
mg}$, and quality factor $Q=760\,000$.

The cavity finesse is $\mathcal{F}=\pi/\gamma=110\,000$, mainly
limited by the losses and transmission of the input mirror. We use
a 500-$\mu$m long cavity in order to keep a sufficient cavity
bandwidth ($\Omega_{\rm cav}/2\pi=1.4\,{\rm MHz}$) and to prevent
laser frequency noise from limiting the displacement sensitivity.
The cavity is operated in vacuum to increase the mechanical
quality factor.

A Ti:Sa laser working at 810 nm provides two cross-polarized beams
used to lock and to probe the cavity. Two acousto-optic modulators
(AO in Fig. \ref{fig:Setup}) enable to detune one beam with respect
to the other. The overall resonance is controlled by locking the
laser frequency via a Pound-Drever-Hall technique: the low-power
($200\,\mu$W) locking beam is phase-modulated at 20\,MHz by a
resonant electro-optical modulator (EO), and the resulting intensity
modulation of the reflected beam provides the error signal. The more
intense ($P^{\rm in}\simeq 4\,{\rm mW}$) probe beam can then be
arbitrarily detuned from the cavity resonance by using the frequency
shift of the AO modulator. A mode cleaner filters the spatial
profile of both beams, while their intensities are stabilized by a
servo-loop which drives the amplitude control of the AO modulators.

The signal $X_{\rm sig}$ is a modulation of the optical cavity
length, obtained through a modulation of the laser frequency
\cite{Caniard} via an electro-optical modulator inserted inside
the laser. The phase $q^{\rm out}$ of the reflected probe beam is
monitored by a homodyne detection, with a local oscillator derived
from the incident beam and phase-locked in order to detect the
phase quadrature.

We first select the detuning of the probe beam with respect to the
cavity resonance, and we monitor the mirror thermal noise by
sending the homodyne detection signal to a spectrum analyzer. This
step allows one to determine the effective mechanical response
$\chi_{\rm eff}$ induced by dynamical back-action. Then, using a
network analyzer, the modulation of the laser frequency is turned
on and swept around the mirror mechanical resonance. The
modulation power is set about $25\,{\rm dB}$ above the thermal
noise at the mechanical resonance so that thermal noise can be
neglected. The resulting phase modulation of the reflected probe
beam is monitored by the network analyzer. We finally turn the
probe beam off, and we measure the mirror thermal noise
immediately after, using the Pound-Drever-Hall signal. This last
step is essential in order to accurately determine the intrinsic
mechanical response $\chi$ of the moving mirror (obtained with the
locking beam at resonance), which may be slightly frequency
shifted from one measurement to the other due to slow thermal
drifts -typically $0.1\,{\rm Hz}$ per minute.

\begin{figure}
\includegraphics[width=6.5 cm]{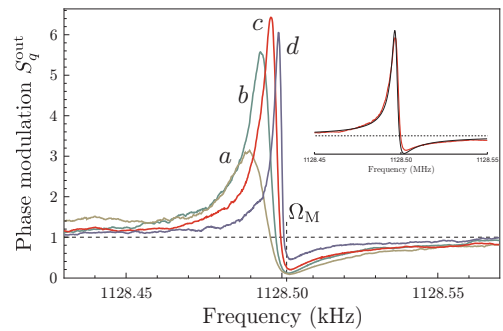}
\caption{Phase modulation power $S_q^{\rm out}$ measured by the
network analyzer in response to a signal modulation swept around the
mechanical resonance frequency $\Omega_{\rm M}$. Curves {\it a} to
{\it d} correspond to different negative detunings,
$\overline{\psi}/\gamma = -1.87,\ -2.03,\ -2.97,$ and $-3.64$,
respectively. The inset shows a fit of curve {\it c} by the
amplification factor $|\chi_{\rm eff}/\chi|^2$ deduced from the
measured susceptibilities $\chi_{\rm eff}$ and $\chi$.}
\label{fig:result1}
\end{figure}

We present in fig. \ref{fig:result1} the resulting phase
modulation power $S_q^{\rm out}$ measured by the network analyzer
when the signal modulation $X_{\rm sig}$ is swept around the
mechanical resonance frequency $\Omega_{\rm M}$. Curves {\it a} to
{\it d}, obtained for various negative detunings
$\overline{\psi}$, are normalized to the phase modulation obtained
far from the mechanical resonance (measured $1\,{\rm kHz}$ above
$\Omega_{\rm M}$). They thus represent the amplification factor
$|\chi_{\rm eff}/\chi|^2$ appearing in eq. (\ref{eq:sensibilite}),
as can be seen from the inset which compares the experimental
result to the expected amplification factor deduced from the
measured susceptibilities $\chi_{\rm eff}$ and $\chi$.

A clear amplification is observed near the effective mechanical
resonance of $\chi_{\rm eff}$, which is down-shifted from
$\Omega_{\rm M}$ as expected from eq. (\ref{eq:chieff}), whereas
one gets an attenuation at the mechanical resonance $\Omega_{\rm
M}$ where $\chi$ is maximum. Note that similar results are
obtained for positive cavity detunings, corresponding to the
amplification regime of the mirror-cavity system rather than to
the cooling one, but the proximity of the parametric instability
\cite{NatureLKB} makes the results less reproducible.
Nevertheless, we have obtained a very large signal amplification
effect, with an amplification factor larger than 6 for curve {\it
c}: back-action effects induce a motion $X_{\rm m}^{({\rm sig})}$
of the mirror in phase with the signal $X_{\rm sig}$, with an
amplitude larger than the signal itself.

Such an amplification leads to an improvement of the sensitivity
beyond the standard quantum limit. Although quantum noises are not
directly observed in our experiment for which the sensitivity is
currently limited by the mirror thermal noise, the sensitivity
improvement can be computed using eq. (\ref{eq:sensibilite}) and
the experimental radiation-pressure amplification results. Since
the resonance frequency $\Omega_{\rm M}$ is on the same order as
the cavity bandwidth $\Omega_{\rm cav}$, finite bandwidth effects
have to be included. This amounts to modify eq.
(\ref{eq:sensibilite}) as follows,
\begin{eqnarray}
\frac{|u|^2 S_q^{\rm out}}{4\xi^2} &=& \left|\hbar \chi_{\rm
eff}\right| \left(\frac{\zeta^{-1}+\zeta}{2} + |v|^2 \frac{\zeta}{2}
+ {\rm Im}\left[v^\star \frac{\chi_{\rm eff}}{\left|\chi_{\rm
eff}\right|}\right]\right) \nonumber\\
&&+ \left|\frac{\chi_{\rm eff}}{\chi}\right|^2 S_x^{\rm
sig},\label{eq:sensibiliteFB}
\end{eqnarray}
where the optomechanical parameter now reads $\zeta=2\hbar \xi^2
|\chi_{\rm eff}|/|u|^2$, and the dimensionless parameters $u$ and
$v$ only depend on optical parameters,
\begin{equation}
u = \frac{\Delta}{\gamma^2 + \overline{\psi}^2 - i\gamma\Omega\tau},
\ v = \frac{\Omega}{\Omega_{\rm cav}} \frac{\gamma
\overline{\psi}}{\Delta} u. \label{eq:uv}
\end{equation}
Radiation-pressure amplification is unaltered by finite-bandwidth
effects [second line in eq. (\ref{eq:sensibiliteFB})], and only
quantum noises are modified (first line), including in particular
the possibility to squeeze the field by radiation-pressure
effects. In the conditions of our experiment, corrections however
do not exceed 1 dB at frequencies for which the amplification
factor is larger than $1$. Large sensitivity improvements can thus
be fully attributed to signal amplification.

\begin{figure}
\includegraphics[width=7.5 cm]{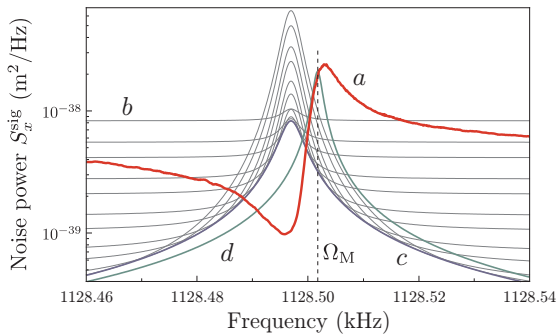}
\caption{Expected quantum-limited sensitivity of the measurement
expressed as the equivalent signal noise spectrum $S_x^{\rm sig}$.
Curve {\it a} is deduced from the signal amplification observed at
a detuning $\overline{\psi}/\gamma = -2.97$. Curves {\it b}
correspond to a resonant cavity with a resonator of susceptibility
$\chi_{\rm eff}$, for increasing powers. Curves {\it c} and {\it
d} are the standard quantum limits for a resonator of
susceptibility $\chi_{\rm eff}$ and $\chi$, respectively.}
\label{fig:result2}
\end{figure}

Figure \ref{fig:result2} shows the sensitivity of the measurement
defined as the equivalent signal noise spectrum $S_x^{\rm
sig}[\Omega]$ that gives a signal-to-noise ratio equal to $1$.
Without radiation-pressure amplification, that is for a resonant
cavity with a mirror of susceptibility $\chi_{\rm eff}$, it simply
corresponds to the quantum noise spectrum given by the first line
in eq. (\ref{eq:sensibiliteFB}). As shown by the series of curves
{\it b}, a flat sensitivity profile is obtained at low input power
where radiation-pressure effects are negligible over the whole
frequency band. As the input power increases, the shot-noise
limited sensitivity is improved away from the mechanical
resonance, at the expense of larger radiation-pressure effects
close to the resonance. Sensitivity is in any case limited by the
standard quantum limit $|\hbar\chi_{\rm eff}|$ shown as curve {\it
c}.

In contrast, the sensitivity in presence of signal amplification
corresponds to the quantum noise spectrum [first line in eq.
(\ref{eq:sensibiliteFB})] divided by the amplification factor
$|\chi_{\rm eff}/\chi|^2$: curve {\it a} shows the resulting
sensitivity, each parameter involved in eq.
(\ref{eq:sensibiliteFB}) being experimentally determined.
Sensitivity is improved beyond the SQL by more than $9\,{\rm dB}$
for frequencies close to the effective mechanical resonance. It is
also improved beyond the SQL $|\hbar\chi|$ of a resonator of
susceptibility $\chi$ (curve {\it d}) by a factor larger than
$5\,{\rm dB}$.

We have experimentally demonstrated how one can use a detuned
cavity to amplify an interferometric signal with the intracavity
radiation pressure, and possibly beat the SQL. Using a fused
silica moving mirror as end mirror of a high-finesse cavity, we
have achieved a 6-fold amplification of the signal by
radiation-pressure back-action. The corresponding quantum-limited
sensitivity would be lower than the SQL for frequencies around the
effective mechanical resonance frequency of the moving mirror. A
similar effect is expected in second-generation GW interferometers
to create a dip in the sensitivity curve and will be used to tune
the sensitivity at a specific frequency.

\vskip 10pt

This work was partially funded by the FP7 Specific Targeted
Research Project Minos.


\end{document}